# Microwave Enhancement of Phase Slip Rate in Quasi One-Dimensional Superconducting Nanowires

Myung-Ho Bae[1,2,*], R.C. Dinsmore III[1], M. Sahu[1], and A. Bezryadin[1]

[1]*Department of Physics, University of Illinois at Urbana-Champaign, Urbana, Illinois 61801-3080, USA*

[2]*Micro and Nanotechnology Lab, Department of Electrical and Computer Engineering, University of Illinois at Urbana-Champaign, Urbana, Illinois 61801-3080, USA*

We study current-voltage (V-I) characteristics of short superconducting nanowires of length ~ 100 nm exposed to microwave radiation of frequencies between 100 MHz and 15 GHz. The radiation causes a decrease of the average switching current of the wire. This suppression of the switching current is modeled assuming that there is one-to-one correspondence between Little's phase slips and the experimentally observed switching events. At some critical power P* of the radiation a dissipative dynamic superconducting state occurs as an extra step on the V-I curve. It is identified as a phase slip center (PSC). With the dependence of the switching currents and the standard deviations observed at the transitions (i) from a constant supercurrent state to a normal state and (ii) from a constant superconducting state to a PSC state, we conclude that both of the two types of the switching events are triggered by the same microscopic event, namely a single Little's phase slip. We show that the Skocpol-Beasely-Tinkham model is not applicable to our microwave-driven phase slip centers, since it leads to an un physical small estimated value of the size of the dissipative core of the PSC. Through the analysis of the witching current distributions at a sufficiently low temperature, we also present evidence that the quantum phase slip play a role in switching events under microwaves.

## I.  INTRODUCTION

Fluctuations, both thermal and quantum, play an important role in determining the physical properties of a one-dimensional superconducting systems. These fluctuations cause the occurrence of events, called Little's phase slips (LPS), in which the phase difference between the ends of the wire (i.e. the phase of the superconducting order parameter) "slips" by $2\pi$ [1]. In other words, the spiral representing the order parameter of a current-carrying state in the Argand diagram, extended with one more axis representing the position along the wire, looses one turn. Note that the number of turns multiplied by $2\pi$ gives the total phase difference between the ends of the wire. At high temperatures, thermal activation of phase slips is the dominant mechanism and is well understood both experimentally and theoretically [2]. At sufficiently low temperatures, quantum phase slips (QPS) are possible [3], but experimental observations of QPS were indirect and remain a topic of active research. The QPS evidence was obtained through observations of an excessive resistance at low temperatures or excessive switching current fluctuations at high bias currents [4]. The final proof of the quantum nature of observed phase slips should come from an observation of a quantum discreteness of the energy levels of the device, or some other inherently quantum effect, which cannot be produced, even in principle, by any sort of electromagnetic (EM) noise or thermal fluctuation. The best known example of such inherent quantum effect is the level anti-crossing [5]. The level anti-crossing occurs because the system, if it is truly quantum, can be either in a symmetric "Schrodinger cat" state (which has a lower energy) or an antisymmetric one (which has a higher energy). In relation to nanowires, it has been suggested that superconducting nanowires (SCNWs) can be used as quantum phase-slip junction [6].

Historically, the definitive experimental observation of a quantum behavior in superconducting devices was first achieved by Martinis, Devoret, and Clarke in the experiments involving microwave (MW) probing of the quantum, i.e. discrete, spectrum of the device [7]. Such approach, involving MW irradiation of the superconducting device under investigation, can also be applied to superconducting nanowires. Therefore, understanding of the MW radiation effect on the phase slips in nanowires is important.

Here we show the general MW response of superconducting nanowires. When a MW signal is applied to the wires and its power is increased, first, we observe a reduction in the critical switching current, $I_{SW}$, followed by the appearance of a dynamical superconducting state, i.e. a phase slip center (PSC). The behaviors of $I_{SW}$ of the wire and of the PSC were intensively studied. It is concluded that the triggering mechanism for different types of switching events (namely, the switching from the constant supercurrent state to the phase slips center and the switching from the constant supercurrent state to the normal state) are triggered by the same underlying microscopic fluctuation effect – the Little's phase slip. Evidence is obtained that quantum tunneling of phase slip is strong at low temperatures (~0.35 K).

In previous studies PSC have been observed in wires near $T_C$ due to the fact the wires were rather thick due to technical limitations. By increasing the temperature it was still possible to achieve one-dimensional (1D) superconductivity regime since the coherence length (which needs to be larger than the wire diameter in order to get 1D regime) diverges near $T_c$. The main difference is that our nanowires are extremely thin and remain quasi-one-dimensional (quasi-1D) down to zero temperature, while most of previous studies [8] have been done on wires which are quasi-1D only close to $T_C$. Thus our type of wires opens up a possibility of studying the dynamics of the quasi-1D condensate, including phase slip centers, down temperatures approaching absolute zero. On the other hand, in our type of suspended nanowires the Joule heating prevented us from observing PSC without MW applied. Therefore a study under MW radiation is carried out and presented here.

## II. FABRICATION AND MEASUREMENTS

As shown in Fig. 1, we fabricate nanowire devices based on the technique of molecular templating [2] originally described in Ref. [9] and further improved in Ref.

[10]. With this technique we are able to make superconducting or insulating single nanowire devices as well as single superconducting wire resonators and devices with multiple wires in parallel. The fabrication is achieved by placing a fluorinated single carbon nanotube over a trench in the substrate and then sputter-coating the nanotube with desired superconducting material.

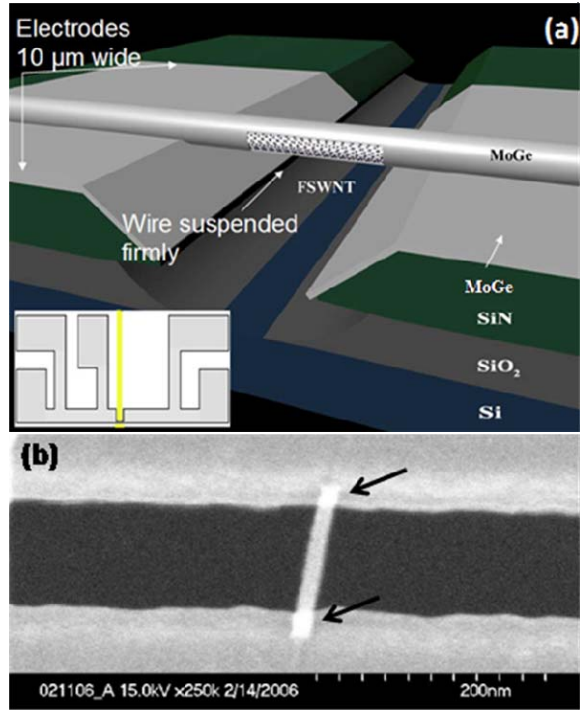

*Fig. 1. MoGe nanowire prepared by molecular templating method. (a) Schematic of the molecular templating technique for the growth of nanometer-scale nanowires. Fluorinated single wall carbon nanotubes are deposited across a trench defined on a substrate and are then sputter coated with a layer of MoGe alloy 5-15nm thick. Insert: Schematic of the contact pads (gray) and the electrodes (gray) connected to the wire (the wire is one the yellow line in the inset). The electrodes are made in the same sputtering run as the wire itself. The yellow line in the inset shows the trench over which the nanotube, which is coated with the metal during the sputtering, is positioned. (b) SEM images of a nanowire. The brighter spots (shown by the arrows) near the banks of the trench indicate that the wire is suspended straightly and firmly across the gap.*

A ~100 nm width and ~5 mm long trench is prepared on a 60 nm thick SiN film on 500 nm thick $SiO_2$/Si wafer by electron beam lithography and reactive ion etching process. To define an undercut in exposed $SiO_2$ layer through the trench, the $SiO_2$ is etched in a ~50 % concentrated HF-in-water solution. This undercut ensures the electrical disconnection between electrodes across the trench except through the suspended nanowire. To form the wire, fluorinated carbon nanotubes are distributed on the substrate from a solution in isopropyl alcohol. After blowing the solution with a nitrogen gas, a desired superconducting metal is deposited by dc magnetron sputtering. Here 5-15 nm of $Mo_{79}Ge_{21}$ (MoGe) is deposited on the suspended nanotubes to form the nanowires. During this process, the metal covers the top of the nanotubes. Figure 1(b) shows a SEM image of a typical suspended nanotube that is coated with MoGe. The beginning and the end of the suspended nanotube show white regions, so-called "white spots", as indicated by arrows. This white spots appear in the SEM images because the nanotube is suspended over the tilted side of the trench (see Fig. 1(a)). Thus the image is formed by the electrons scattered by the tube and by the side of the trench. They indicate that the nanowire is straight, which is important to exclude the possibility of formation of weak link in the superconducting nanowire.

All samples are wired in a pseudo-four-probe configuration (Fig.1(a), insert), with each pair of voltage and current lines a twisted pair of measurement lines. The bias current is applied by using a precision voltage function generator Stanford Research Systems (SRS) DS360 connected in series with a standard resistor $R_s$. The $R_s$ value is much larger than the sample resistance. The voltage across the standard resistor and the voltage across the sample are amplified using either an SRS SR530 or a PAR 113 preamplifier. The preamplifiers are battery powered and thus have a low noise level at the input terminals. The outputs of these preamps are fed to an analog-digital convertor board. The temperature is measured using a Ruthenium Oxide thermometer which is wired in a 4-probe configuration and measured using a Lakeshore 370A Temperature controller. To achieve the desired temperature the output current of the temperature controller is either applied to a heater attached to the the 3He pot or the sorption pump heaters. Using this device we are able to maintain temperatures within approximately 5 mK or better. The microwave signal is generated by a Gigatronics 1026 function generator. This device can output signals from -99 dBm to + 10 dBm for frequencies in the range of 1 MHz- 26 GHz. The signal is capacitively (and inductively) coupled to the sample through an antenna positioned at the bottom of the sample can. Because of this we were only able to study the sample's response at resonant frequencies of the Faraday cage, in which the MW antenna and the sample with a nanowire are located.

## III. RESULTS AND DISCUSSION

### A. Microwave response on superconducting nanowires

Figure 2(a) shows current-voltage characteristics (*V-I*) of the nanowires at various temperatures, which show a hysteretic behavior below *T*=2.9 K. At T=358 mK, as bias current is swept from zero to higher values, the wire switches from a superconducting state to a resistive or normal state due initially to an LPS and subsequent to Joule heating associated with the wire switching to the normal state (which is observed as a large voltage jump on the V-I curve). It was found previously that the switching is triggered by each single LPS at low temperatures, while at higher temperatures, while at higher temperatures (empirically, higher than ~1 K) a temporal approximate coincidence of a few LPS is needed to overheat the wire and produce the switching [4,11].

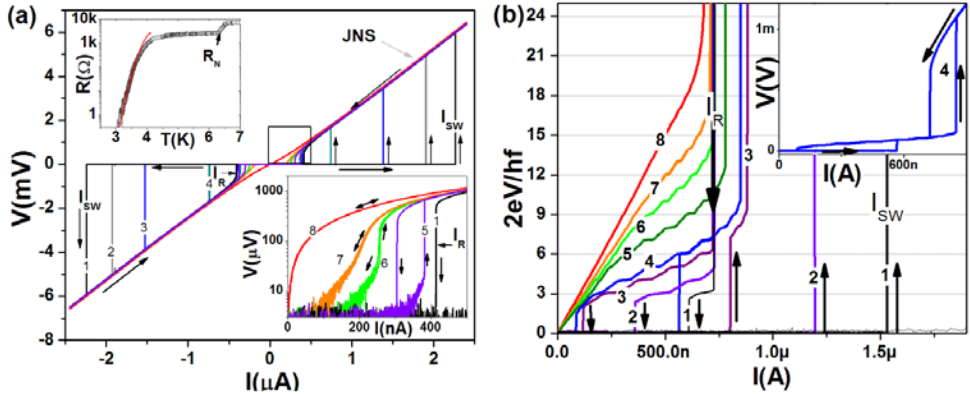

Fig. 2. (a) V-I characteristics for sample 121707C numbered according to increasing temperature (no microwave signal is applied here): (1) T = 358mK; (2) T = 1K; (3) T = 1.8K; (4) T = 2.5K; (5) T = 2.9K; (6) T = 3.1K; (7) T = 3.4K and (8) T = 4.1K, above $T_C$ of the wire. The critical currents $I_{SW}$ and $I_R$ are indicated. After the switching at $I_{SW}$ the wire reaches the normal state resistance. Lower Inset: log scale of the boxed region with curves 2-4 removed for clarity. The curves 6, 7, 8 are measured at higher temperatures, at which the switching events are smeared due to thermal fluctuations. Upper Inset: R vs. T for the sample with fit to TAPS theory with parameters $R_N$=2760 Ω, $T_C$=4.1K, L= 110nm, ξ(0)=10.2nm. The resistance drop at 6.5 K is due to the leads going superconducting. The normal resistance of the wire $R_N$ is taken to be the value measured right below the temperature at which leads become superconducting. This value is indicated by the arrow. (b) Positive bias V-I characteristics for sample 121707C as the microwave signal of various powers is applied. The frequency is f=7.34 GHz and the temperature is T=350 mK. The voltage drop on the wire is normalized by the radiation photon energy, as indicated (here h- Planck's constant, e- electron charge and f is the frequency). Shapiro steps are clearly observed at integer values of the normalized voltage. The curves are numbered according to increasing microwave power, P, measured at the source output, in dBm: (1) -5.1; (2) -4.7; (3)-4.2; (4) -3.4; (5) -2.5; (6) -1; (7) 0.2 and (8) 2. Inset: the general view of the V-I curve under microwaves. The formation of a phase slip center, helped by the applied microwave radiation, is observed at low bias as a resistive state with the resistance much lower than the normal resistance.

The corresponding switching current $I_{SW}$ is stochastic, meaning that each time the *V-I* curve is measured one obtains a slightly different value for $I_{SW}$. Once in the normal state, the *V-I* curve is linear with the slope almost exactly equal to the normal resistance of the entire wire. The normal state is marked "JNS" (standing for Joule normal state) in Figure 2(a). In the insert of Fig. 2(b) the JNS is also visible at the highest current bias shown in the plot. As the bias current is decreased considerably below the switching current, the wire jumps back to the superconducting state (ScS) at the retrapping current $I_R$ (see the lower inset of Fig. 2(a)). The superconducting state, ScS is the state in which the voltage on the wire is zero although the current is not zero. This retrapping current is not stochastic, meaning that in every measurement the same value of $I_R$ (with the precision of the setup) is observed (also see Fig. 3(b)). This deterministic nature indicates that the resistive state is simply a normal state, stabilized due to Joule heating of the wire [12]. The same conclusion is confirmed by the fact that the measured resistance in JNS is the same as the normal-state resistance of the wire $R_N$. In this case the retrapping is explained by cooling of the wire below its critical temperature $T_C$ as the bias current is decreased. Since the cooling process involves a macroscopic number of degrees of freedom, associated with normal electrons, the fluctuation of the $I_R$ is not observed (the relative value of fluctuations is proportional to $1/\sqrt{N}$ where N is the number of degrees of freedom involved).

Increasing the temperature from 358 mK suppresses $I_{SW}$ while the retrapping current remains constant up to $T\sim 2.5$ K $< Tc$ (Fig.3(a)). At intermediate temperatures 2.5 K $< T <$ u3.1 K both $I_{SW}$ and $I_R$ are suppressed with increasing temperature and there is a finite voltage due to thermally activated phase slips which appears as "resistive tails" in the *V-I* curves, before the switching occurs [4] (e.g. see the curve 5 in the lower inset of Fig.2 (a)). This fact is important evidence indicating that at high enough temperatures single phase slips can occur in the wire without always necessarily causing the switching into the Joule-heated normal state (JNS). In other words, many phase slips are required to occur almost simultaneously to generate enough heat to switch the wire into JNS. Such multiple phase slip process is well described by the models developed by Shah et al. [11]. The same model, and the absence of the resistive tails at low temperatures (roughly below 1 K) indicate that at low temperatures each single phase slip cause the observable switch, either into the JNS (or into the PCS state if microwaves are applied) [4,11]. As the temperature increases from $T = 3.1$ K to $T = T_C (= 4.1$ K), the hysteresis disappears and the critical current goes to zero.

The Fig. 2(b) shows the MW response (for frequency $f = 7.34$ GHz) on $V$-$I$ curves at $T = 350$ mK. The MW response of the wire shows some similarity to the temperature response but it is different enough to rule out the trivial heating of the nanowire with the MW radiation. When the power of the MW signal increases from zero, $I_{SW}$ of the wire decreases and $I_R$ does not change. In that respect the effect of MW is like the effect of heating. At higher powers of applied MW, an additional resistive branch occurs in the $V$-$I$ curve [see Fig. 2(b) inset], either before the switching to the normal state (as the current is swept up, so the corresponding branch is called the sweeping-up branch) or after the retrapping from the normal state (as the current is swept down, corresponding to the sweeping-down branch). The occurrence of the branch takes place at a certain power, $P^*$, in the sweeping-up branch, and at $P^R$ in the sweeping-down branch of the $V$-$I$ curve. We interpret this new resistive branch (which is not observed if the temperature is increased and no MW is applied) as a MW-assisted phase slip center (PSC). This observation broadens the options of investigating the dynamics of the order parameter in nanowires since without the radiation it is impossible to observe a PSC, since the wire always overheats and jumps into the JNS. These observations are analogous to the observations of Anderson and Dayem on thin film constrictions [13], performed on rather wide superconducting bridges. The effects of increasing MW power on the critical currents are better summarized in Figure 3. As the microwave power increases from 0 to $P^*$ the switching current $I_{SW}$ for ScS→JNS transition monotonically decreases while remaining stochastic and the JNS→ScS transition at $I_R$ remains constant and deterministic. Figure 3a shows this trend by plotting 100 values for $I_{SW}$ and $I_R$ as a function of MW power. The kink in the plot is indicative of the onset of a PSC branch in the sweeping-up branch, which occurs at the power, which is denoted as $P^*$.

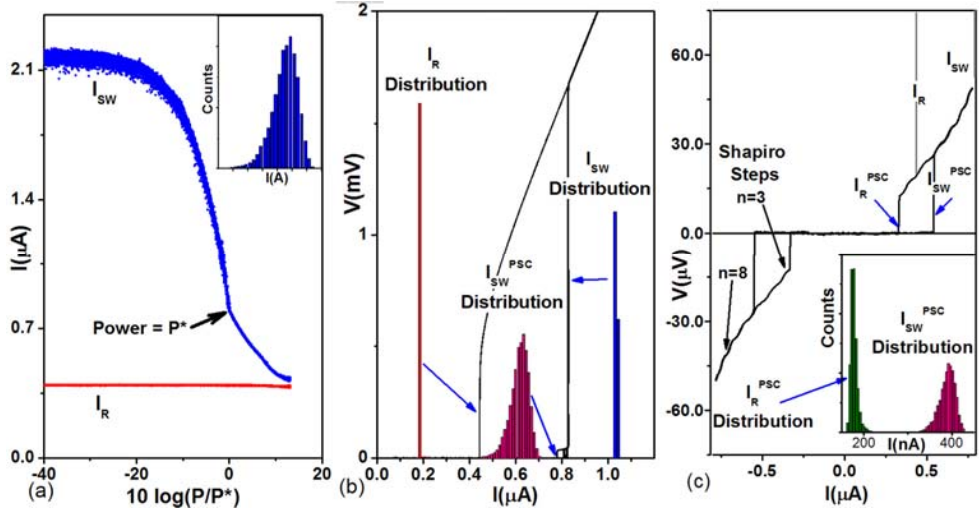

*Fig. 3. (a) Switching and retrapping currents versus microwave (MW) power at 3 GHz for sample 121707C. The switching current $I_{SW}$ corresponds to the switching into the normal state, and retrapping from the normal state, $I_R$. As many as 100 points were taken at each power showing the spread due to the stochasticity of $I_{SW}$ at low microwave powers. At all powers for $I_R$ and at high powers of MW the spread in $I_{SW}$ is smaller than the point size. The power P\*, indicated above, is defined as the power when a PSC first appears in the switching branch (Fig.2b-insert). The power P\* also corresponds to the kink in the curve. Inset A representative histogram for $I_{SW}$ at power P < P\*. (b) Positive bias segment of a V-I curve for 10 traces for a power just above P\*. The histograms shown are for counts of a given $I_R$, $I_{SW}$ and $I_{SW}^{PSC}$. The jumps the statistics of which is measured are indicated by the blue arrows. The probability distributions are computed from 10,000 total scans. The distributions are scaled to fit the figure. Only the switching current from the superconducting state to the phase slip center ($I_{SW}^{PSC}$) shows a broad distribution. (c) V-I curve at P > P\* and at a bias slightly below V-I. Shapiro steps are visible with steps corresponding to V=nhf/2e with n=3 and n=8 indicated by arrows. Inset: distributions for the switching current, $I_{SW}^{PSC}$, and the retrapping current, $I_R^{PSC}$, for the PSC for 10,000 scans. Both critical currents for the PSC are stochastic. All data shown here were measured at a temperature of 500 mK.*

For powers higher than P\* two jumps in V-I curve are observed as the current is increased: the first jump represents ScS→PSC and the second jump corresponds to PSC→JNS, (Figure 2(b), inset). Thus, for P > P\*, we introduce two notations for the jumps: $I_{SW}^{PCS}$ representing the transition ScS→PSC and $I_{SW}$ representing the transition PSC→JNS. Thus in all case the symbol $I_{SW}$ represents the current at which the wire switches to the normal state (either from a completely superconducting state or from a PSC state). Histograms in Fig. 3(b) shows that while the transition ScS→PSC remains highly stochastic, the transition PSC→JNS exhibits a very narrow, almost deterministic distribution. As the current is reduced we also observe either one or two retrapping events. The retrapping current for the transition *from the normal branch* to either the fully superconducting branch or the PSC state is denoted as $I_R$. The retrapping current for the transition from the PSC branch to the fully superconducting branch is denoted as $I_R^{PCS}$.

The PSC is a dynamic superconducting state as evidenced by the appearance of Shapiro steps and the stochastic nature of both the switching current ($I_{SW}^{PSC}$) and retrapping current ($I_R^{PSC}$) (see Fig. 3(c)). Unlike the ScS↔JNS transition, the ScS↔PSC transition is clearly stochastic in both directions.

The transition ScS→PSC follows the same trend as the transition ScS→JNS at powers lower than $P^*$, in the sense that (a) the width of the distribution of $I_{SW}$ for $P < P^*$ is very similar to the width of the distribution of $I_{SW}^{PSC}$ at $P > P^*$, and (b) the slope of $I_{SW}$ vs. $P$ for $P < P^*$ is very similar to the slope of the curve $I_{SW}^{PCS}$ vs. $P$ at $P > P^*$ (see Fig. 5). These observations strongly indicate that the same physics for the transition out of the ScS to the JNS and out of the ScS to the PSC is involved. The Shapiro steps occur due to synchronization of the superconducting phase difference rotation with the applied high frequency MW radiation and were discussed in detail in [14]. The steps appear at voltage values given by *2eV=nhf*, as expected, where *e* is the elementary charge, *n* is the integer number, and *h* is the Plank constant. Shapiro steps are only present in the portion of the *V-I* curve attributed to the PSC. The lack of Shapiro step in the high voltage regime (i.e the JNS regime) independently verifies that it is a non-coherent (i.e. normal) state.

**B. Microwave-induced phase slip centers and the Skocpol-Beasely-Tinkham (SBT) model**

Our quasi-1D wires are too narrow to support a normal vortex core even at zero temperature thus enabling us to study PSCc down to temperatures much lower than the critical temperature. In the Skocpol, Beasely and Tinkham (SBT) model [15] of the PSCs the order parameter is suppressed in a small region of the wire that is on the order of $2\xi$ in length. The generated Bogoliubov quasiparticles (bogoliubons) [16] then diffuse a distance $\Lambda_Q$, a length longer than $\xi$ near $T_C$, away from the center of the PSC before relaxing back into the condensate.

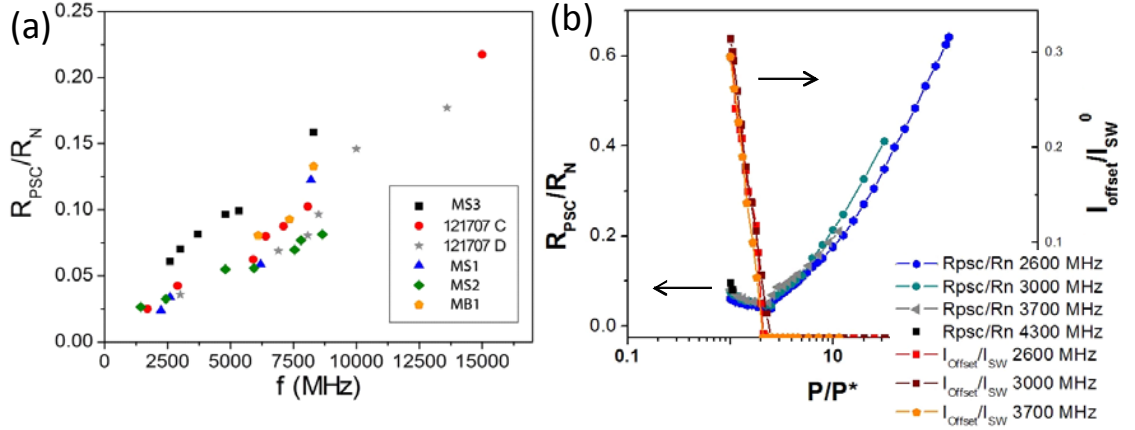

*Fig. 4. (a) Average resistance (normalized by $R_N$) versus frequency for six samples taken at ~350mK and MW power slightly greater than P\*. The data show that the average resistance of the MW-assisted PSC monotonically increases with the frequency of applied MW. (b) $R_{PSC}/R_N$ and $I_{offset}/I_{SW}$ vs. P/P\* for three frequencies for the sample MS3. Resistance is taken as a linear fit of the first few steps in the V-I characteristic. A kink in $R_{PSC}$ occurs near where the offset current goes to zero.*

This model predicts the voltage of the PSC to be given by: $V=2\rho\Lambda_Q(I-\beta I_S)/A$ with *I*, the DC bias current, *A* is the cross sectional area of the wire, $\rho$ is the resistivity of the wire material (typically $\rho$~200 $\mu\Omega$ cm), $\Lambda_Q$ is the bogoliubon diffusion length, and $\beta$, a constant on the order of ½ such that $\beta I_S$ is the offset current, which is the average supercurrent in the PSC (far from the PCS we have $I=I_S$). This model predicts the differential resistance of the PSC to be given by $R_{PSC}=2\rho\Lambda_Q/A$, or equivalently $R_{PSC}/R_N=2\Lambda_Q/L$ with $R_N$ and *L* being the normal state resistance and the length of the nanowire. The differential resistance we measure in MW-assisted PSC depends strongly on the MW frequency and ranges from nearly zero to 1 k$\Omega$ at radiation powers slightly above P\*. Figure 4 is a plot of $R_{PSC}/R_N$ vs *f* for six different samples. Note that according to the SBT model, if applied to our observations, the ratio $R_{PSC}/R_N$ should be equal to the length of the dissipative region (which is the length over which the quasiparticles propagate before relaxing into the condensate) of the PSC divided by the total length of the wire. That the length of our nanowires is about 100 nm and the coherence length, $\xi$ is about 10 nm. So, according to SBT, the minimum expected value for the ratio $R_{PSC}/R_N=2\Lambda_Q/L$ is 0.2, because the quasiparticles should at least exist within the core, which cannot be smaller than the coherence length. Yet smaller values have been observed, as low as 0.025 for this ratio (Fig.4a). The observed very low values of this ratio, suggesting according to the SBT model that the core is smaller than the coherence length in some cases, appear to be unphysical. Thus we conclude that the SBT model is not applicable to our case involving low temperatures and the MW radiation. Hopefully the future theoretical work will lead to an appropriate generalization of the SBT model to describe MW-assisted PSCs, described in this work.

We attempt to use this model here because (a) no theory for MW-induced PSC does exist at present and (b) to emphasize the qualitatively different physics observed on MW-induced and sustained PSC compared to regular PSC. The average resistance of the PSC ($R_{PSC}$) was taken by a linear fit of the $V$-$I$, in the range $I_{SW}^{PSC} < I < I_{SW}$. The range of the fit is such that a few Shapiro steps are covered. We find a monotonic dependence of thus defined average PSC resistance on MW frequency (Fig.4a). This linear dependence might be related to the fact that the height of the steps is proportional to the frequency, through the well-known formula $V_n \sim hfn/2e$. We are not aware of any theoretical model applicable to this result.

Figure 4(b) shows measurements of $R_{PSC}/R_N$ and the normalized offset current, $I_{offset}/I_{SW}$, for sample MS3 as a function of applied MW power at three different frequencies. It is interesting to note that the resistance of the PSC measured in this manner decreases slightly for increasing power and then has a discontinuous jump with the applied MW power. On the other hand, the offset current decreases steadily to zero as the MW power is increased.

The presence of the PSC branch can be explained as follows: If the system jumps to the PSC state at a low enough bias current then overheating might be small and the wire would not jump to the JNS immediately, but only when the current is strong enough to generate heating (at which point the second jump, now to the normal sate, is observed). The offset current value determines where the PSC branch intersects the overheating-controlled hysteresis curve. The offset current for the PSC decreases with increasing microwave power so a low resistance PSC will appear first in the current-sweeping-up branch at $P^*$ and then in the current-sweeping-down branch when the offset current, which decreases with applied power, is decreased below $I_R$. At higher frequencies the average slope of the PSC branch becomes higher and for powers below $P^*$ the intersection of the PSC branch with the overheating transition occurs at a very high voltage and the power dissipated in the wire, P=$IV$, is high enough to heat the wire above $T_C$. Thus, when the wire switches into the PSC branch it immediately goes normal because the wire cannot dissipate the heat generated by the bias current. In this case the PSC appears in the retrapping branch at $P^R$ which is the power at which the offset current is sufficiently below $I_R$ so that the PSC branch is stable. A PSC that appears first in the current-sweeping-down branch will appear in the current-sweeping-up branch at a slightly higher power. At higher power the switching current, $I_{SW}$, is further reduced and it intersects the PSC branch at a lower voltage. This voltage is low enough that the Joule heating is not sufficient to heat the wire above $T_C$. At powers in this regime a stable dynamic superconducting state exists and the system stays in the PSC branch until the current is increased to $I_{SW}$.

## C. Transition into the normal state

The transition current from PSC to JNS at $I_{SW}$ is not stochastic and therefore its dynamics is different from that of the transition from ScS to the PSC. This transition, PSC→JNS, is in some ways similar to the switching transition at higher temperatures near $T_C$ in the absence of MWs: these transitions show very strong phase diffusion in the form of "voltage tail" occurring before the switching transition (see the curve 5 in the bottom inset of Fig. 2(a)). The voltage before the switching to the normal state can be seen and compared in curves 5 in Fig. 2(a) and curves 4 and 5 in Fig. 2(b). It has been shown that the phase diffusion can be enhanced by MW radiation in superconductor-insulator-superconductor (SIS) Josephson junctions having $\hbar I_{C0}/2e >> E_C$ [17] where $E_C$ is the Coulomb charging energy of the junction and $I_{C0}$ is its critical current. For the samples considered, using the maximum switching current as an estimate for the critical current, $I_{C0}$, one gets $E_J{\sim}0.5$ meV, whereas the Coulomb energy, $E_C$, given by $e^2/(2C)$, using $C = 1\ fF$, is about 10 μeV [18]. Thus our system is in the regime where this model predicts that MWs cause the enhancement of phase diffusion. The authors in Ref. [17] predict that the maximum supercurrent that an SIS junction can carry when MWs are present scales as $P^{1/4}$. To test if this mode is applicable to our system we analyzed data for various observed switching transitions in several wires. Figure 5 shows the transition currents ($I_{SW}$) plotted vs. applied MW power for sample 121707C taken at 2.9 GHz. The higher power region for $I_{SW}$ before it merges with the $I_R$ was fit to determine its power dependence at several frequencies to determine the exponent $\alpha$ in the relation: $I_{SW} \propto P^{-\alpha}$.

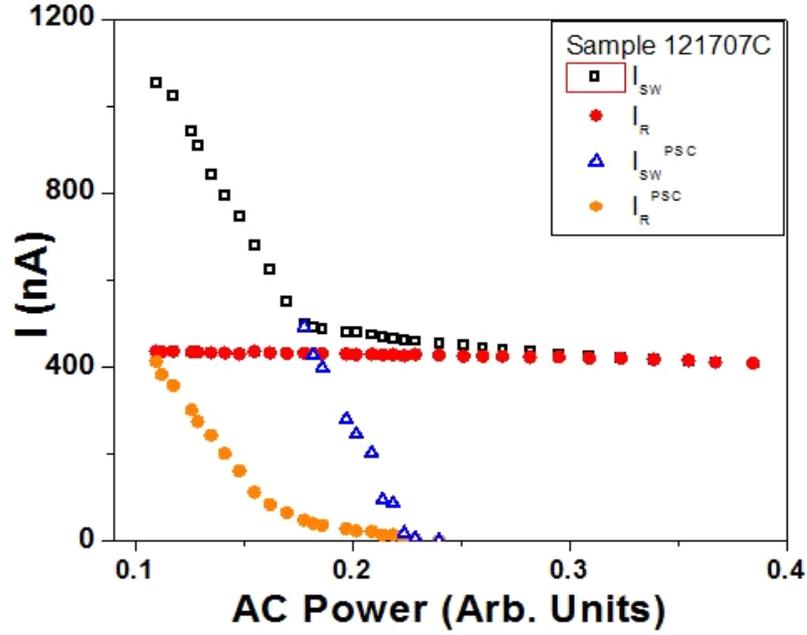

*Fig. 5. Mean values of various switching and retrapping currents in sample 121707C at 2900 MHz plotted vs. applied MW power. Note how $I_{SW}^{PSC}$ exhibits the same slope as $I_{SW}$ once the critical power is reached at which there are two switching currents. This indicates that the same triggering mechanism must underlay the transition ScS→JNS (for P<P\*) and ScS→PSC (for P>P\*). This triggering mechanism is identified as a single LPS (for low temperatures, i.e. T < 1 K).*

Figure 6 summarizes the results on sample 121707C. The data were taken by varying the MW power and measuring the switching current 100 times at each power. The curve taken at 3 GHz is the same one that appears in Fig. 3(a). The results of these fits are summarized in Table 1 for three samples measured at different frequencies.

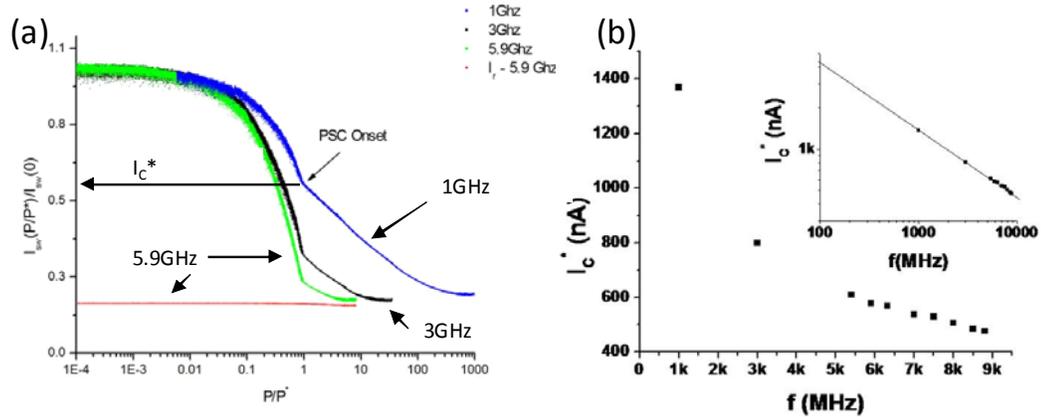

*Fig. 6. Switching current behavior for different frequencies for sample 121707C. (a) $I_{SW}$ vs. applied microwave power, P for three frequencies as shown. $I_C^*$ is indicated for the curve taken at 1 GHz. The retrapping current $I_R$ is shown for 5.9 GHz and shows the same behavior for other frequencies. The portion of the curve for powers greater than $P^*$ are fit to determine the power dependence $I_{SW} \propto P^{-\alpha}$. (b) $I_C^*$ plotted vs. the frequency. Inset: log-log plot of same data. The slope of a linear fit was -0.479.*

| Sample | α in $I_{SW} \sim P^{-\alpha}$ | MW frequency (GHz) |
|--------|------------------------|--------------------|
| 121707C | 0.19 | 1 |
| 121707C | 0.25 | 5.4 |
| 121707C | 0.2462 | 5.9 |
| 121707C | 0.2217 | 8.5 |
| 121707C | 0.21 | 8.8 |
| MS3 | 0.17 | 3 |
| MB3 | 0.24 | 7.3 |

**Table 1** *Summary of fits of switching current $I_{SW}$ vs. Power with Figure 6(a). The results are close to the predictions of the Koval-Fistul-Ustinov model [17].*

Although the experimental values of α varied significantly from sample to sample and from one frequency to another, this variations might be within the range of uncertainties and the noise of the experiment. The powers are all normalized to $P^*$ for comparison. The curves showed the same trend at different frequencies, however their shapes are different and the current at which the PSC appears (i.e. the power $P^*$) showed some frequency dependence. The frequency dependence is shown in Figure 6(b). When plotted on a log-log scale the frequency dependence looks very regular and further theoretical analysis is needed to understand this behavior.

### D. Switching Current Distributions and Quantum Phase Slips

Before discussing the behavior of the switching event in a wire exposed to MW radiation, we will first review the recent results on this process in the purely DC case without MW [4,11]. In Ref. [4], the authors studied the stochastic nature of the switching current at different temperatures to understand the underlying mechanism that leads to the transition from the superconducting state to the JNS. What they found was that the distribution of switching currents became broader as the temperature was lowered and that this rather counter-intuitive trend is due to a crossover of the multiple phase slip switching mechanism, occurring at high temperatures, to a single-phase slip switching mechanism (when each phase slip causes a switch to JNS with necessity), taking place at lower temperatures. In order to understand quantitatively the occurrence of the switching events and their relationship with phase slips, they model the thermal conductivity of the superconducting nanowire system [11]. The wire is a suspended structure measured in a vacuum. So the heat dissipated in the wire in the form of bogoliubons must be conducted to the leads where it can be dissipated further. Joule heat is generated when a phase slip occurs. When this happens a small region of the wire becomes heated and possibly normal. Thus it will generate more heat, due to Joule heating effect, since the wire is biased with a constant current. The amount of heat dissipated by each phase slip is $Ih/2e$ [11], where $I$ is the bias current. Using this relationship and a numerical model for thermal conductivity it was shown that at temperatures lower than approximately 1 K a single Little's phase slip is sufficient to cause the wire to overheat and switch from the superconducting state to the JNS.

To determine whether this single phase slip is due to thermal or quantum fluctuations the switching current distribution (SCD) can be converted into a switching rate using the Kurkijarvi method [19], which was extensively used to analyze switching events in current-biased SIS junctions in Ref. [20] (the model is developed for the cases when the bias, which reduces the barrier for the switching, is increased linearly in time). The measured rates are then compared to the rates for the two types of phase slip processes, namely thermally activated phase slips (TAPS) (LAMH model) and quantum phase slips (QPS) (Giordano model) [21,22]:

$$\Gamma_{\text{TAPS}} = \Omega_{\text{TAPS}} \exp\left(-\frac{\Delta F(T,I)}{k_{\text{B}}T}\right)$$

$$= \left(\frac{L}{\xi(T)}\right)\left(\frac{1}{\tau_{\text{GL}}}\right)\left(\frac{\Delta F(T)}{k_{\text{B}}T}\right)^{1/2} \exp\left(-\frac{\Delta F(T,I)}{k_{\text{B}}T}\right) \qquad (1a)$$

$$\Gamma_{\text{QPS}} = \Omega_{\text{QPS}} \exp\left(-\frac{\Delta F(T,I)}{k_{\text{B}}T_{\text{QPS}}}\right)$$

$$= \left(\frac{L}{\xi(T)}\right)\left(\frac{1}{\tau_{\text{GL}}}\right)\left(\frac{\Delta F(T)}{k_{\text{B}}T_{\text{QPS}}}\right)^{1/2} \exp\left(-\frac{\Delta F(T,I)}{k_{\text{B}}T_{\text{QPS}}}\right) \qquad (1b)$$

With $\tau_{\text{GL}} = [\pi\hbar/8k_{\text{B}}(T_{\text{C}}-T)]$ the Ginzburg-Landau relaxation time, $\Omega$, the attempt rate, $T_{QPS}$, the temperature dependent effective temperature for a quantum phase slip form the Giordano model, having the free energy barrier given by [12,23]:

$$\Delta F(T,I) = \frac{\sqrt{6}\hbar I_{\text{C}}(T)}{2e}\left(1-\frac{I}{I_{\text{C}}}\right)^{5/4} \qquad (2)$$

Their results [4] corresponding to low temperatures (0.3 - 1 K) were in good agreement with the Giordano's model of QPS and were not consistent with the TAPS model. Thus switching events statistics at low temperatures was used to obtain evidence for QPS.

### E. The Effect of Microwaves on the Switching Current Distribution

We have already discussed how MWs can lead to the suppression of the switching current, but now we will discuss how they affect the SCD. Figure 7 shows SCDs and the corresponding switching rates versus the bias current for the sample MS3 for increasing applied MW power from -44 dBm to -17 dBm. According to the overheating model as introduced in the previous section [11], this wire is in the single QPS regime at 400mK (and also below this temperature). Thus a single phase slip (either thermal or quantum) is always sufficient to cause the wire to switch from a superconducting state to the JNS state. In this case the rate of QPS equals the experimentally measured switching rate, so the Eq. 1(b) can be applied. However we need to incorporate the effect of the applied MW radiation into this model. When we expose the wire to the MW radiation we introduce an ac bias to the wire in addition to the dc bias. In this case we can compute the free energy barrier for a phase slip to occur using the following equation:

$$\Delta F(T,I) = \frac{\sqrt{6}\hbar I_{\mathrm{C}}(T)}{2e}\left(1 - \frac{I_{DC} + I_{RF}\sin\omega t}{I_{\mathrm{C}}}\right)^{5/4} \qquad (3)$$

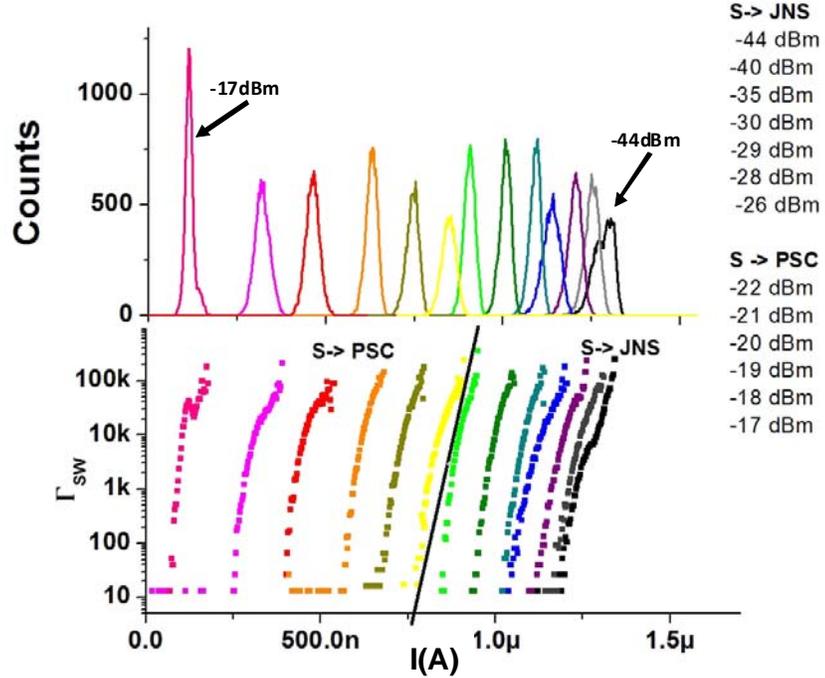

Fig. 7. Top: Switching current distributions (SCD) and the corresponding rates for the sample MS3 at different MW powers taken at T=400mK. Top: SCD for the powers shown in the legend. The SCD shifts to the left for higher powers. The SCDs were obtained by measuring the first voltage jump, as the bias current is increased. At low power this jump reaches the normal state while at higher MW power the first jump leads to the phase slip center (PSC) creation. Bottom: switching rates computed from the above

*distributions, using the Kurkijarvi theory [19,20]. The black line denotes the division between the transitions from Superconducting state to JNS and to the PSC. Note that there is no significant difference between the two types of transitions indicating that a PSC state may always be the state to which the wire switches first, at least for a short time. In other words, our explanation to the absence of any qualitative difference between the curves to the left and to the right of the black line is the following: Even if the observable switch leads to the normal state, the initial jump happens to the superconducting dynamic PSC state, which then quickly overheats. This data was taken using a 50 Hz sweep to measure the V-I curve.*

, where $I_{DC}$ is the dc bias current and $I_{RF}$ is the ac bias current with angular frequency, $\omega$, induced in the wire from the MW antenna and $t$ is the time. Using Eqs. 1(b) and 2, we can compute the switching rate for a single QPS event. To do this we first average over one period of the MW signal to get the time averaged switching rate (since the setup used to measure the switching rate is much slower than the frequency of the applied microwave). Using this strategy we can compute the switching rate as a function of temperature, DC bias current, $I_C(T)$, $\xi$, $T_C$, $T_{QPS}$ and the AC bias current $I_{RF}$ and use this computed curve to fit the experimental data. The parameters $\xi$ and $T_C$ are taken to be the values obtained using TAPS fits and $T_{QPS}$ and $I_C(T)$ are determined by the overheating model for temperature dependence of the switching current.

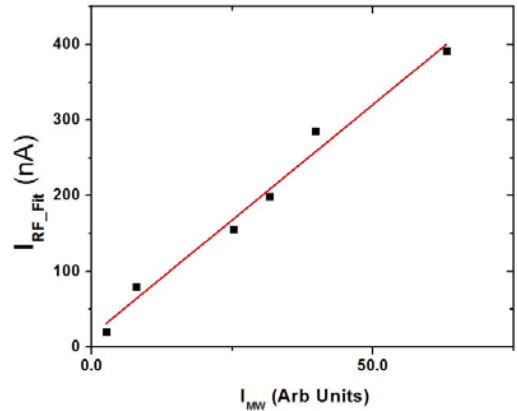

*Fig. 8. Fitting parameters vs. applied MW current for sample MS3. The rate data are those between - 40dBm and -26dBm shown in Figure 7. For these fits $I_C(T=0K)$ was 1.6 μA and the fits started to fail to fit the data for $I_{RF} \sim 600$ nA. The rates for higher power are for the PSC regime which may have a different switching behavior.*

These parameters may be changed slightly from those values however they remain fixed for all rates at different MW powers, so the only parameter that is changed to fit the rates measured at various applied MW powers is $I_{RF}^{Fit}$. SCDs are measured for one wire at a constant temperature at several MW powers. The rates are then computed using the above procedure and they are fit using this model by varying $I_{RF}^{Fit}$. The absolute value of the experimental MW induced amplitude $I_{MW}$ is not known, however it must be proportional to the square root of the applied MW power. This applied ac bias current, $I_{MW}$, in arbitrary units, is computed from the applied MW power as $I_{MW}=(10^{P/10})^{1/2}$, where $P$ is the power in dBm. If the fits done with this model on a set of data produce $I_{RF}$ values that are proportional to $I_{MW}$ then the model is consistent with the experiment. It is important to note that this simple model is not expected to work for values of $I_{RF}$ that approach closely the critical current. Figure 8 shows the results of applying the fitting procedure to the low power rate data for the data shown in Figure 7.

Figure 9 shows σ, which is the standard deviation (i.e. the characteristic width) of the distributions of the type shown in Fig.7, as a function of the applied MW power. From this graph we see that σ for the S→JNS transition is similar to its value for the S→PSC transition.

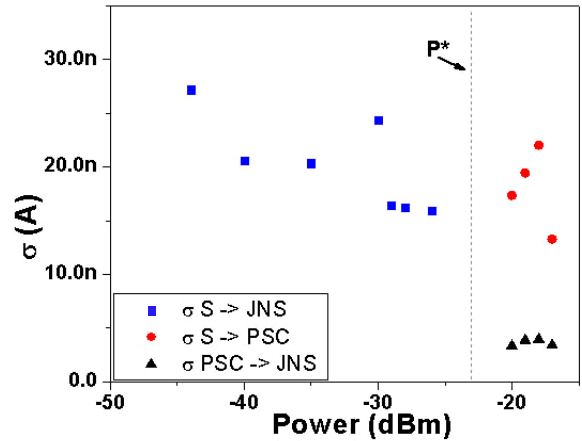

*Fig. 9. Standard deviation of SCD vs. applied MW power for sample MS3. P\* is denoted by the dashed line. We observe that σ for the transition S→JNS for P < P\* is similar to σ for the transition S→PSC for P > P\*: both are about 20 nA. For the transition PSC→JNS σ is about 4 nA.*

This also leads us to conclude that the transition out of the superconducting state is governed by the same process for each type of transition, namely by the initial nucleation of a phase slip center (this term is used to denote a spot where the phase slips periodically, not just one time) through the occurrence of a single LPS first.

The results were confirmed by measuring another sample, 091608B. Figure 10 shows SCD and the computed switching rates for this sample taken with a 5 Hz sweep rate and 10,000 points for each distribution. The rates (Fig.10, bottom panel) are fit to the single QPS model using the procedure outlined above. The results of the fitting procedure for the values of $I_{RF}^{Fit}$ (which is a fitting parameter in the model representing the amplitude of the MW induced ac current in the wire) are plotted vs. the applied dimensionless microwave current, $I_{MW}$ in Figure 11.

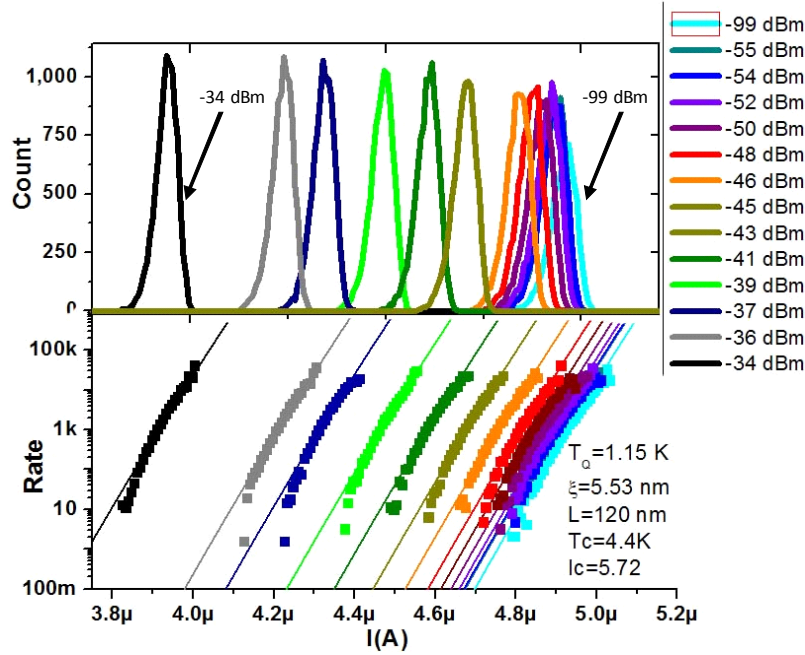

*Fig. 10. SCDs and rates at different MW powers taken at T = 300 mK for sample 091608B. Top: 10,000 point SCDs taken at the MW powers given in the legend. Bottom: Rates for the SCDs in the top graph along with fits to the single QPS model. The fitting parameters are shown in the lower legend and are close to the values used in the overheating model. The values for $I_{RF}^{Fit}$ (and only this parameter was changed) were changed to fit each curve. The resulting values of the fitting parameter are shown in Fig 11.*

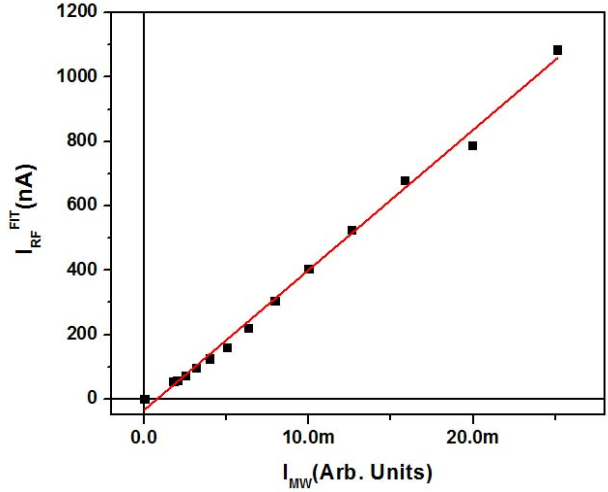

*Fig. 11. Fitting parameters, $I_{RF}^{Fit}$, vs. applied MW current for sample 091608B. The other parameters used are those given in Figure 10. The linear fit shows that our fitting parameters and the fitting model, expressed by Eq.(3), is consistent with the experiment.*

The linear fit in Fig. 11 is rather good, indicating that the proposed interpretation based on the assumption that single QPS events trigger the observable jumps on the *V-I* curve (the first jump, to be more precise) under MW irradiation.

**F. Unexplained Phenomena at Low Microwave Frequencies**

We also find that PSCs appear at low frequencies, ~100 MHz without observable Shapiro steps and with the differential resistance on the PSC step of the order of 10 Ω , which is much lower than the normal resistance of the wire or even the resistance of a wire segment of length ξ.

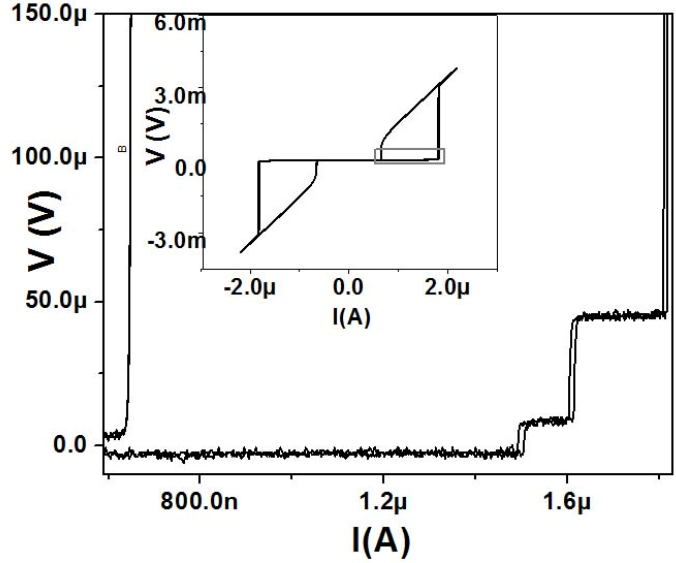

*Figure 13. V-I for sample MS2 with MW at 100 MHz and T= 350 mK. These steps are termed "Low-frequency steps" or LFS. They appear to be different from Shapiro steps. Inset: Full scale plot of the same V-I curve.*

Figure 13 shows an example *V-I* curve for the sample MS2 when it is exposed to lwo frequency MW radiation (100-200 MHz). The main difference between these steps and the steps visible in, e.g., Fig. 2(b) (measured at a high MW frequency) is that they do not seem to correspond to the phase lock-in effect. The steps of Fig. 13 can be called "low-frequency steps (LFS)". The differential resistance observed within each LFS is close to zero. The steps also have a very large negative offset current. These LFS do not follow the behavior predicted by the SBT model and thus remain unexplained.

Another phenomenon that was only observed at lower frequency was the appearance of zero-crossing voltage steps in sample MB2. Zero-crossing means that there is a finite voltage on the wire at zero dc bias applied bias current. These results were similar to what we observed in focus ion beam milled BSCCO Intrinsic Josephson Junction (IJJ) samples [24]. The phenomenon of zero-crossing is not well understood in any superconducting system. Figure 14 shows measurements of this phenomenon on sample MB2. A nanowire in exhibiting this behavior could be used as a rectifier since an ac current induced by a microwave is essentially converted to a dc voltage. This behavior is also interesting because it may provide evidence for an effective ratchet potential for phase slips in our wires under certain conditions.

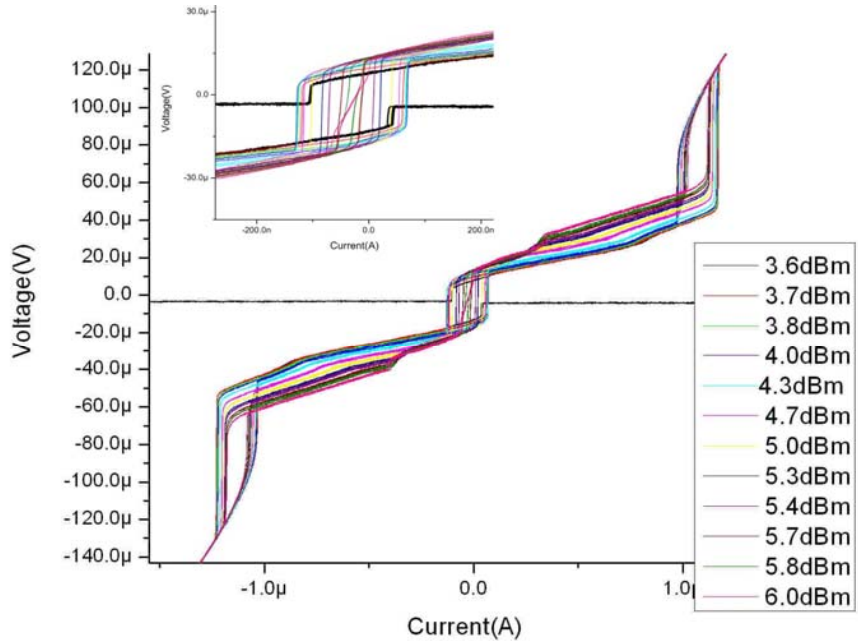

**Figure 14.** *Zero-crossing voltage steps in sample MB2 for different MW powers at 320 MHz. Inset shows zoomed in region to highlight the zero-crossing. The shape of these curves are very similar to those observed on the intrinsic BSCCO junctions [24].*

## IV. CONCLUDIONS

The physical properties of MW induced and sustained PSCs were examined in detail. We show that the PSCs observed in the *suspended* superconducting nanowires under MWs do not behave like those studied near $T_C$ in previous experiments. We compare the switching transitions ScS→JNS, occurring at zero or low powers, to the switching transitions of the type ScS→PSC at sufficiently high MW powers. We present strong evidence that those transitions are each caused by the same triggering mechanism, namely the LPS. Further, we compare the results with the Giordano model for the quantum phase slip rate. We present evidence that single quantum phase slips represent the main cause for the switching events at low temperatures even under MWs.

## ACKNOWLEDGEMENTS

This work was supported by the DOE grant DE-FG02-07ER46453.

*E-mail address: mhbae@illinois.edu

**REFERENCES**

---